\documentclass[twocolumn,showpacs,amsmath,pre]{revtex4}
\usepackage{graphicx}

\begin{document}
\title{Local mean-field study of capillary condensation in
silica aerogels}

\author{F. Detcheverry, E. Kierlik, M. L. Rosinberg, and G. Tarjus}
\affiliation{Laboratoire de Physique  Th{\'e}orique des Liquides,  Universit{\'e} Pierre et
Marie Curie, 4 place Jussieu, 75252 Paris Cedex 05, France}

\date{\today}
 
\begin{abstract}
We apply local mean-field (i.e. density functional) theory to  a lattice model of a
fluid in contact with a dilute, disordered gel network. The gel
structure is described by a diffusion-limited cluster
aggregation model. We focus on the influence of porosity on both the
hysteretic and the equilibrium behavior of the fluid as one varies the
chemical potential at low temperature. We show that the shape of the
hysteresis loop changes from smooth to rectangular as the porosity
increases and that this change is associated to disorder-induced out-of-equilibrium
phase transitions that differ on adsorption and on desorption. Our
results provide insight in the behavior of $^4$He in silica aerogels. 

\end{abstract}
\pacs{64.60.-i,68.45.Da,75.60.Ej}
\maketitle

\def\be{\begin{equation}}
\def\ee{\end{equation}}
\def\bea{\begin{eqnarray}}
\def\eea{\end{eqnarray}}

\section{Introduction}

The influence of quenched disorder on phase transitions and critical phenomena continues to be the focus
of intensive experimental and theoretical research activity. Major effects are expected and actually 
 observed when the disorder couples linearly to the order parameter of the 
system, a situation that is realized when a fluid or a fluid mixture is confined within a porous glass or 
is in contact with the interconnected strands of a gel.  A dilute
rigid network  is a particularly interesting medium from a theoretical perspective because exclusion (i.e., confinement) effects do not play a dominant role, so that  the random-field Ising model (RFIM) may be a useful framework to interpret the experimental observations\cite{BG1983}. 

A striking example of the influence of a gel network on fluid phase behavior
is provided by the thermodynamic studies of Chan and co-workers
on $^4$He in silica aerogels of varying porosity. Silica
aerogels are  highly porous, fractal-like solids  made of a tenuous
network of SiO$_2$ strands interconnected at random sites. In a $95\%$ porosity aerogel, specific
heat and adsorption (vapor-pressure isotherm) measurements\cite{WC1990}  performed
in the vicinity of the critical temperature of the pure fluid
($T_c$=5.195K) show evidence of a phase separation between a low-density  ``vapor'' phase 
(presumably composed of $^4$He vapor plus a liquid film around the silica strands\cite{LMPMMC2000}) and a high-density  ``liquid'' phase filling the whole pore space.  The first-order transition appears to 
terminate at a sharply defined critical point that is only 31mK below
$T_c$, which suggests that the system is in a weak random-field
regime. The coexistence boundary in the presence of
aerogel is, however, much  narrower than in the pure system. Subsequently, similar results were obtained
with N$_2$ in the same aerogel, using light scattering and vapor-pressure isotherms\cite{WKGC1993}.
Since  out-of-equilibrium and hysteretic effects due to domain
formation are generally observed in random-field systems below the
critical temperature of the pure system (as illustrated by the
behavior of binary mixtures in aerogels\cite{BFC1997} and  diluted
antiferromagnets in a magnetic field\cite{B1998}), it is noteworthy
that no hysteresis  is present in the measurements performed in
Refs.\cite{WC1990,WKGC1993}.  The gel-fluid system  has thus reached
equilibrium within the time scale of the experiments, which is
indeed found  much longer than the characteristic time of activated dynamics\cite{WKGC1993}. 
The situation changes, however, at lower temperatures and the adsorption of  $^4$He in a $98\%$ porosity aerogel  is clearly hysteretic at 3.60K and 2.34K\cite{C1996,TYC1999}. Both adsorption and desorption isotherms  display  a vertical step at well-defined pressures, but draining occurs  at a lower pressure than filling\cite{note2}. 
The shape of the hysteresis loop, moreover, depends  on the porosity of 
the aerogel: in an aerogel of 87\% porosity, $^4$He adsorbs and desorbs gradually at 2.42K and there is 
no signature of a ``liquid-vapor'' phase coexistence (see Fig. 4(c) in
Ref.\cite{TYC1999}).  Such hysteretic behavior is reminiscent of capillary
condensation in a low-porosity solid like Vycor glass  where one
observes a rapid increase of the adsorbed quantity at a pressure below
the bulk saturated vapor pressure, but no sharp vertical step in
the adsorption isotherms\cite{E1967}.  The mechanism for hysteresis in
porous substrates has prompted much discussion in the literature and
various explanations have been proposed,  focusing either on
single-pore metastability {\it  {\`a} la} van der Waals or on network pore-blocking effects\cite{BE1989}. 
Since both models seem completely inadequate to describe light
aerogels, the $^4$He experiments raise several 
questions: what is the scenario for the change in the shape
of the hysteresis loops\cite{note3}? Do filling and draining obey different  mechanisms?  What is the true equilibrium behavior when hysteresis is present?

These questions are addressed in the present work where we build on
our earlier studies of capillary condensation in disordered
porous solids\cite{K2001,K2002,R2003}, but focusing on aerogels. We are mainly
interested in the influence of the porosity on the hysteretic behavior of  sorption isotherms. This is partly
motivated by theoretical studies of the zero-temperature
RFIM which predict the existence of a disorder-induced out-of-equilibrium
phase transition in the hysteresis loop\cite{S1993}. The
first experimental observations of this phase transition have been
recently reported in the literature\cite{B2000}, and we argue that the
change in the $^4$He adsorption isotherm from sharp to smooth can be
interpreted within the same framework. We propose a different scenario
for draining and we relate the observed behavior to an
out-of-equilibrium phase transition 
associated to the depinning of the liquid-vapor interface.
The approach we develop is similar
to that used previously for calculating the irreversible behavior of spin
glasses, random-field ferromagnets and diluted
antiferromagnets\cite{SLG1983}; it consists in studying the evolution of
the free-energy surface (more precisely, the grand-potential surface
$\Omega$) as the external driving field (here, the pressure $P$ of the
external vapor or, equivalently, the chemical potential $\mu$) is changed. 
$\Omega$ is a functional of the local fluid
density and is treated in the mean-field approximation. At low
temperatures this  multidimensional free-energy  landscape is characterized by a large
number of local minima in which the system may  be trapped. The main physical assumption underlying our
description is that  thermally activated processes  play a negligible role on the time scale 
of the experiments (in other words, the dynamics is similar to that at $T=0$). 
The evolution of the system then proceeds either continuously by the
deformation of the local minimum in which the system is trapped or
when this latter becomes unstable by a jump to another minimum
(avalanche); the response to the driving field is then discontinuous and irreversible.  Since it is impossible to perform
such a study for a  continuous model because of computational limitations
(especially when finite-size scaling analysis is required near phase transitions), we
adopt a lattice-gas description that incorporates at a coarse-grained
level the geometric and energetic disorder of the gel-fluid mixture\cite{KKRT2001}.
Previous work has shown that many of
the phenomena observed in experiments on fluids in
disordered porous solids  can be reproduced qualitatively by
such simple lattice models\cite{K2001,K2002,R2003,WSM2001}. As in other
studies of phase transitions in aerogel\cite{UHR1995,STC1999}, we
model the solid by a
 fractal structure obtained by diffusion-limited
cluster-cluster aggregation. 
  
The paper is organized as follows. In Sec. II, we introduce the
lattice model and discuss some important features of the aerogel structure. In Sec. III, we 
present the local mean-field theory and describe the numerical
procedure. The results for the adsorption, desorption, and equilibrium
isotherms in $87\%$ and $95\%$ porosity aerogels at $T/T_c=0.45$ are given in Sec. IV. The final section presents a summary and conclusions.

\begin{figure}[h]
\begin{center}
\resizebox{8cm}{!}{\includegraphics{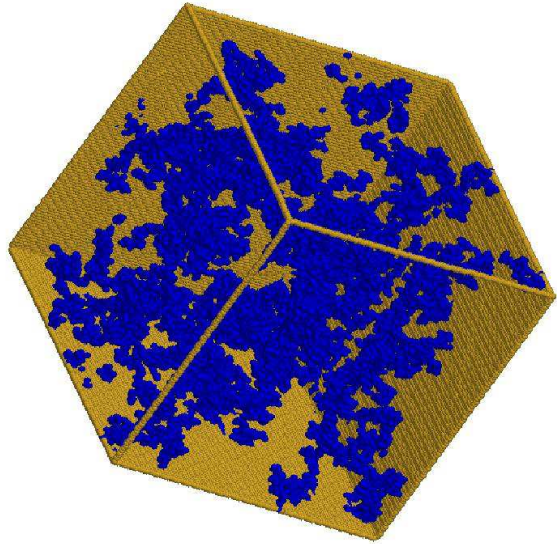}}
\caption{Three-dimensional realization of  a 98\% porosity DLCA aerogel on a 
bcc lattice of size $L=100$ with periodic boundary conditions.}
\end{center}
\end{figure}

\section{Lattice model and aerogel structure}

 The lattice model used in this work describes the solid  as a
collection of fixed impurities that exert a random yet correlated  (by
the connectivity of the strands) external field on the atoms of the
fluid. The Hamiltonian is given by\cite{PRST1995,KRTP1998} 

\bea
{\cal H} = &-&w_{ff}\sum_{<ij>} \tau_{i}\tau_{j} \eta_i \eta_j  -\mu \sum_i \tau_i
\eta_i\nonumber \\
&-&w_{sf}\sum_{<ij>
}[\tau_{i}\eta_i (1-\eta_j)+\tau_{j}\eta_j (1-\eta_i)]
\eea
where  $\tau_i=0,1$ is the usual fluid occupation variable ($i=1...N$)
and $\eta_i=1,0$ is a quenched random variable that characterizes the
presence of gel particles on the lattice (when $\eta_i=0$, site $i$
is  occupied by the gel); $w_{ff}$ and $w_{sf}$ denote, respectively, the fluid-fluid and
solid-fluid attractive interactions, $\mu $ is the fluid chemical
potential, and the double summations run over all distinct pairs of
nearest-neighbor (n.n.) sites.  One can thus vary
the gel porosity, $p=(1/N) \sum_i \eta_i$, by changing the number of solid sites or modify the ``wettability''  of the solid surface
by  changing the ratio $y=w_{sf}/w_{ff}$. For $y=1/2$, 
the model reduces to a site
diluted Ising model, as can be seen by transforming
the fluid occupation variable $\tau_i$ to an Ising spin variable,
$s_i=2\tau_i-1$\cite{KRTP1998}: preferential adsorption of the liquid
phase onto the gel is thus modeled by $y>1/2$.  Random fields are
generated in the system when $y \neq 1/2$, and the fluctuating part of the
field acting on spin $s_i$ is proportional to the number of solid
particles that occupy  the nearest neighbors of site
$i$\cite{KRTP1998}.   This a discrete random variable that can take
the values $0,1,...c$, where $c$ is the
coordination number of the lattice, and whose probability distribution is strongly porosity-dependent.

Gel configurations (i.e., sets $\{\eta_i\}$) are generated by a standard on-lattice diffusion-limited
cluster-cluster aggregation (DLCA) algorithm\cite{M1983} adapted to  a  body-centered cubic  (bcc) lattice
with periodic boundary conditions. The choice of the bcc lattice will
be  motivated in Sec. IV.B.
The DLCA algorithm mimics the growth process
of base catalyzed  aerogels used in helium experiments, and it has
been shown to reproduce  the main structural features of these
aerogels measured from scattering experiments\cite{H1994}. 
A typical exemple of a $98\%$ porosity DLCA aerogel on a lattice of linear size
$L=100$ is shown in Fig. 1 (from now on, we take the lattice
constant $a$ as the unit length; the total number of sites in the lattice is thus
$N=2 L^3$). One can clearly see
the fractal-like  character of the gel network that results from the aggregation mechanism.

\begin{figure}[t]
\begin{center}
\resizebox{8cm}{!}{\includegraphics{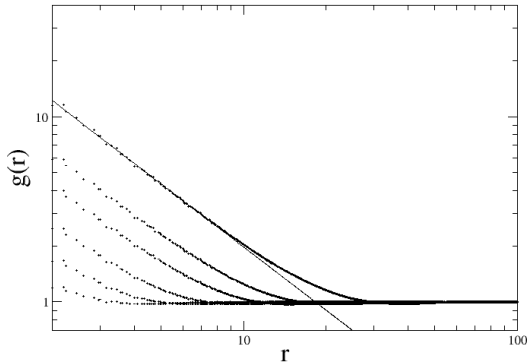}}
\caption{Log-log plot of the aerogel correlation function $g(r)$ for
$p=0.99, 0.98, 0.97, 0.95, 0.92$ and $0.87$ (for top to
bottom). The solid line is a fit with slope $-1.13$ that 
corresponds to the fractal regime for $p=0.99$.}
\end{center}
\end{figure}

More quantitative information on the structural properties of the gel
can be extracted from the two-point correlation function $g(r)$. Each
curve shown in Fig. 2 corresponds to a different porosity $p$ and
results from an average over several simulations in a box of size $L=100$
(for $p=0.87, 0.92, 0.95$) or  $L=200$ (for $p=0.97, 0.98, 0.99$).
Only the most dilute samples exhibit a true intermediate  regime
described by  the power-law behavior $g(r) \simeq r^{-(3-d_f)}$ 
revealing  the fractal character of the intra-cluster density-density
correlations\cite{note4}. For $p=0.99$, one  finds  $d_f\approx 1.87 $, a
value that  lies in the range  $1.7-1.9$  expected for DLCA in three
dimensions. Note, however, that  the fractal dimension decreases with increasing
porosity\cite{L1998}, so that the asymptotic value must be somewhat
smaller. Another estimation of $d_f$ can be obtained from the average
cluster size $\xi_G$ by assuming that $\xi_G$ varies as
$\rho_G^{-1/(3-d_f)}$, where $\rho_G=1-p$ is the gel concentration. As
suggested in Ref.\cite{H1994}, $\xi_G$ can be estimated from the location of the
first minimum of $g(r)$ (which is hardly visible on the scale of
Fig. 2). The plot of $\xi_G$ versus $\rho_G$ shown in Fig. 3 yields $d_f\approx 1.85$. 
Since the correlation length is
in the range $650-1300\AA$ for a $98\%$ base-catalyzed
aerogel\cite{LMPMMC2000,LKMP2000} and $\xi_G(\rho_G=0.02)\approx 21$ in the simulation, one can
estimate that the lattice constant $a$ corresponds to about $30-60\AA$. This is consistent with the 
coarse-grained picture of a gel site representing  a Si0$_2$ particle
with a diameter of about $30\AA$.

Another relevant length scale is the size of the largest
cavity in the aerogel\cite{PP1999}. Fig. 4 shows the distribution $P(n)$ of nearest distances $n$
from an empty site to the aerogel as a function of porosity (the integer
``distance'' $n$ is defined here as the length of the shortest path on the
lattice from an empty site to
a gel site; $n=1$ means that the empty site is n.n. of a gel site). The
plot is the normalized histogram of these distances and 
 $\sum_1^nP(n')$  gives the probability of being closer than $n$ to the
aerogel. One can see that the size  of the largest cavity is $n\approx 32$
for a $99\%$ aerogel and decreases to $n\approx 10$ for a
$95\%$ aerogel and to $n\approx 5$ for a
$87\%$ aerogel. It is noticeable that the distribution changes
significantly when decreasing the porosity from $95\%$ to $87\%$. In the former case,
$P(n)$ has its maximum at $n=2$ and there is a significant proportion of empty
sites that are not in the immediate vicinity of the gel. In the latter
case, $P(n)$ is monotonic and strongly peaked at $n=1$, which
indicates that most of the empty sites are very close to the gel. We shall see in section IV.A that these differences 
have important implications for the behavior of the fluid during adsorption.

\begin{figure}[t]
\begin{center}
\resizebox{8cm}{!}{\includegraphics{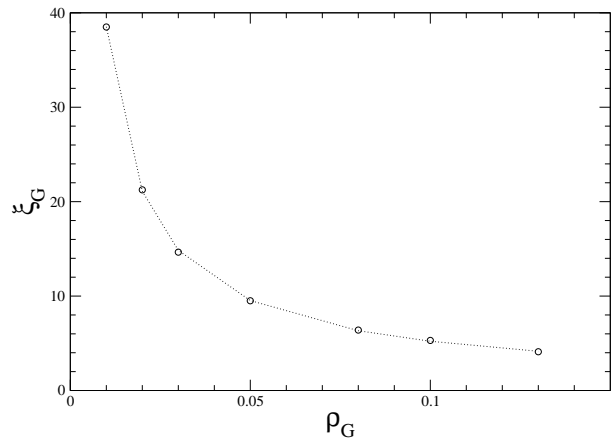}}
\caption{Average cluster size $\xi_G$ (estimated from the location
of the first minimum of $g(r)$) as a function of the gel concentration
$\rho_G=1-p$. The dashed line corresponds to the power law fit $\xi_G=0.707
\rho_G^{-0.868}$.}
\end{center}
\end{figure}

\begin{figure}
\begin{center}
\resizebox{8cm}{!}{\includegraphics{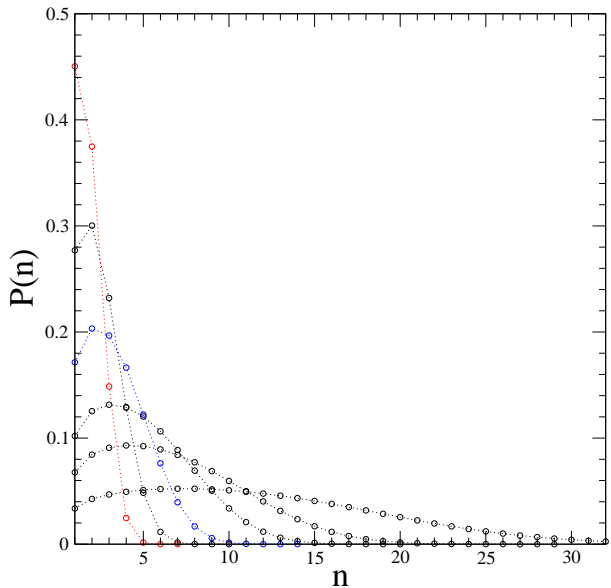}}
\caption{Distribution $P(n)$ of nearest distances from an empty site to
the gel (see
text). From right to left:
$p=0.99, 0.98, 0.97, 0.95, 0.92,0.87$ (the dashed lines are guides
for the eye).}
\end{center}
\end{figure}

It is clear from Fig. 3 that the minimum system size necessary to describe 
correctly  collective effects occuring inside the aerogel on long  length scales (like a sharp condensation event in
the {\it whole} pore space) depends strongly on the porosity. For instance, 
a box of linear size $L=100$ is not large enough to represent the
whole pore space of a $99\%$ aerogel because it
does not contain a sufficient number of connected fractal aggregates
(a problem that cannot be cured by  merely increasing the number of realizations).  Within 
the framework of local mean-field theory, it is however 
impossible to simulate much larger lattices  because the computational 
time and the memory storage requirement become
prohibitive (at finite temperature, the local fluid densities are continuous variables, which forbids 
the use of bits algorithms).
In the present work, we consider $95\%$ and $87\%$ aerogels
for which we can investigate the statistics of collective events
while working with lattices of reasonable size (between $L=25$ and $L=100$).

Since  it has been shown previously that the interface between the porous solid and the bulk gas 
may have  a dramatic  effect on the desorption process\cite{R2003},  two different setups are considered.  In the first one, the system is periodically replicated in all directions so that no interfacial effects are taken into account. In the second setup an interface is created by placing  a slab of  vapor  of width  $L_b=10$ (the gas ``reservoir'') in
contact with one of the $[100]$ faces of the simulation box.  Periodic boundary conditions
are then imposed  in the $y$ and $z$ directions and reflective boundary
conditions in the $x$ direction (obtained by reflecting the lattice at the boundaries). 

In order to completely specify the model, one must also fix the value of the interaction
parameter  $y$. As shown previously\cite{R2003}, the shape of the
hysteresis loops changes with  $y$. Since  it is
meaningless  in a coarse-grained picture to compute $y$ from the actual values of the  fluid-fluid and solid-fluid 
van der Waals interactions, we have chosen its value 
so as to reproduce approximately the height of the hysteresis loop in the $87\%$ aerogel at low temperature.
Specifically, the results presented  in this work are calculated with $y=2$, a value 
for which the filling of the $87\%$ aerogel at the lower closure point of the
hysteresis loop is  roughly the same as in the experiment at $T=2.42$K
(i.e., at $T/T_c \approx 0.45$). This corresponds to about $1/3$ of the
filling reached
at the plateau just before saturation, as can be seen in Fig. 4(c)
of  Ref.\cite{TYC1999} (see also Fig. 5(a) below).  All calculations
are done at this single reduced temperature (recall that
$kT_c/w_{ff}=c/4=2$ in the mean-field approximation).

\section{Local mean-field theory (LMFT) and numerical procedure}

For the present model, the LMFT consists in solving 
the self-consistent  equations for the thermally averaged fluid densities $\rho_i(\{\eta_i\})=<\tau_i\eta_i>_T$
obtained from minimizing the mean-field grand-potential functional\cite{K2001}

\bea
\Omega(\{\rho_i\})&=&k_BT \sum_i[\rho_i\ln \rho_i+(\eta_i-\rho_i)\ln(\eta_i-\rho_i)]\nonumber\\
&-&w_{ff} \sum_{<ij>}\rho_i\rho_j -\mu\sum_i\rho_i\nonumber\\
&-&w_{sf}\sum_{<ij>}[\rho_i(1-\eta_j)+\rho_j(1-\eta_i)] \ .
\eea
The  variational procedure $\delta \Omega/\delta \rho_i=0$ gives a set of  $N$ coupled non-linear equations
\be
\rho_i=\frac{\eta_i}{[1+e^{-\beta (\mu+w_{ff}\sum_{j/i}[\rho_j+y(1-\eta_j)]) }]}\ ,
\ee
where $\beta=1/(k_BT)$ and the sum runs over the $c$ nearest neighbors of
site $i$. At low temperature, these equations may have several
solutions, 
and  the grand potential corresponding to solution $\{\rho_i^{\alpha}\}$ is given by\cite
{K2002}
\be
\Omega^{\alpha}=k_BT \sum_i \eta_i\ln(1-\frac{\rho_i^{\alpha}}{\eta_i})+w_{ff} \sum_{<ij>}\rho_i^{\alpha}\rho_j^{\alpha}  \ .
\ee
By using an iterative method to solve Eqs. (3), one only finds
solutions that are only local minima of the grand-potential surface, i.e., metastable states\cite{SLG1983}.

For a given realization of the aerogel, the sorption isotherms  (i.e.,
the curves $\rho_f=(1/N) \sum_i \rho_i(\{\eta_i\},\mu)$) are obtained by increasing
or decreasing the chemical potential in small steps $\delta \mu$. At each
subsequent $\mu$, the converged solution at $\mu-\delta\mu$ (on adsorption) or at
$\mu+\delta\mu$  (on desorption) is used to start the iterations.  The
isotherms are then averaged  over a number of gel realizations depending on the system size.
In order to determine the equilibrium isotherms, we have also searched for additional solutions of Eqs. (3) inside the hysteresis loop by performing scanning trajectories, as explained below
in section IV.C.

To accelerate the convergence of the numerical procedure, the
iterations have been updated (i.e., the new value of $\rho_i$ is substituted into the r.h.s. of Eqs. (3)
before waiting for the rotation through the indices $i$ to be
complete) and the evolution of each  site density $ \rho_i$ has been monitored in the following way. Fluid sites are divided into two categories, active and passive. 
At the beginning
of the calculation, all sites are active and the procedure stops when all sites are passive. At  each 
iteration only active sites are considered and their new density is
calculated using Eqs. (3).  If
$\mid\rho_i^{(n)}-\rho_i^{(n-1)}\mid/\rho_i^{(n-1)}<10^{-6}$, where $n$ denotes the
$n$th iteration, the site becomes
passive; otherwise its remains active and its nearest-neighors become (or remain)
active (some  sites  can thus be passive during a few iterations and become active
again). 
\begin{figure}[t]
\begin{center}
\resizebox{8cm}{!}{\includegraphics{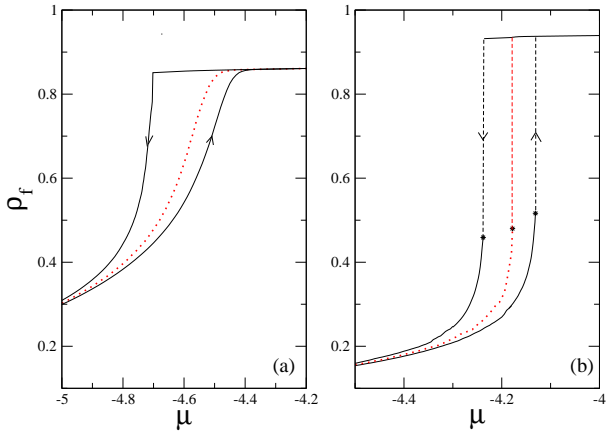}}
\caption{Hysteresis loops in  the $87\%$ (a) and $95\%$ (b) model aerogels
calculated at $T/T_c=0.45$. Equilibrium isotherms are indicated by the dotted lines.}
\end{center}
\end{figure}
The advantage of this algorithm
is that the number of active sites may quickly become  a small  fraction of the
total number of sites, which of course significantly reduces the overall
computation time. For instance, the number of active sites is only
$O(L^2)$ when a planar liquid-vapor interface propagates through the
system. The  algorithm is useful in 
dilute aerogels because of the presence of large cavities. 
We have checked that the numerical results are not different (to the precision of the calculations) from those obtained without updating or with using a more standard convergence criterion, like in our previous studies of fluid adsorption in random matrices\cite{K2001,K2002,R2003}.
Convergence (to an accurracy of $10^{-6}$) typically requires between $10^{2}$ to $10^{3}$
iterations,  and several CPU hours on a $2.4$ GHz workstation 
are needed to calculate a single adsorption isotherm in a system
of linear size $L=100$  (using a step
$\delta\mu/w_{ff}=10^{-2}$ or $10^{-3}$ to increment the chemical potential). The
search for equilibrium isotherms is much  more time consuming (see
below in Sec. IV.C) and has been made possible by using parallel
computation on a Beowulf cluster of $24$ processors.

\section{Results and discussion}

The results of the present study are summarized in Fig. 5 that shows
the adsorption, desorption, and equilibrium isotherms calculated for 
the $87\%$ and $95\%$ model aerogels for $y=2$ and $T/T_c=0.45$ (from now on, $\mu$ is in units of $w_{ff}$, the fluid-fluid interaction parameter).
 These curves result from the detailed  numerical analysis  described
in the following and correspond to the thermodynamic
limit. The
most striking feature is the change in the shape of the hysteresis
loop from smooth to rectangular as the porosity increases. This change
is similar to that observed experimentally by Chan and
co-workers (see Figs. 4(b) and 4(c) in Ref.\cite{TYC1999}). As is explained below, 
we predict that the three isotherms (adsorption, desorption, and
equilibrium) are discontinuous in the $95\%$ aerogel (at least within mean-field theory where thermal
fluctuations are neglected). The underlying 
physical mechanisms are, however, quite different.

\subsection{Adsorption }

\begin{figure}[t]
\begin{center}
\resizebox{8cm}{!}{\includegraphics{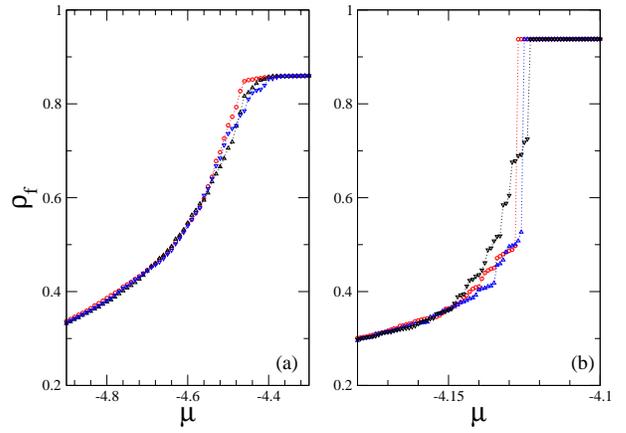}}
\caption{Representative adsorption isotherms in $87\%$ (a) and $95\%$ (b)
aerogels. System sizes are $L=50$ (a) and  $L=100$ (b). Notice the 
change in the scale of the $x$ axis  between the two figures.}
\end{center}
\end{figure}

We first examine the adsorption process. Calculations performed in the
presence and in the absence of a gel/reservoir interface yield isotherms that are
almost indistinguishable (except for the smallest system sizes),
confirming the conclusion reached in our previous work\cite{R2003} that adsorption does not depend on the existence of a free surface. This is also in line with the results of previous
calculations done in single slit-like or cylindrical pores\cite{MVS1989}. 

In Figs. 6(a) and 6(b) are shown some
adsorption isotherms calculated in $87\%$ and $95\%$ porosity samples of linear
size $L=50$ and $L=100$, respectively (this corresponds to roughly the
same ratio $L/\xi_G \approx 10$). In both cases, we focus on  the region where 
the adsorption is the steepest (see Fig. 5). At lower coverages, the isotherms look gradual and 
smooth in  the two systems, but as $\mu$ increases, they consist of
little steps of varying sizes. In the $87\%$ aerogel, the size of these
``avalanches'' remains small all the way up to the slowly increasing plateau that extends to
saturation. On the other hand, in the $95\%$ aerogel, the size of the
jumps tends to increase with $\mu$ and in most of the samples there is a  large final avalanche after which the filling is almost complete (there are however some rare
realizations where the last jump is not much  larger that the preceeding ones, as illustrated by one of the isotherms in Fig. 6(b)).

\begin{figure}[t]
\begin{center}
\resizebox{6cm}{!}{\includegraphics{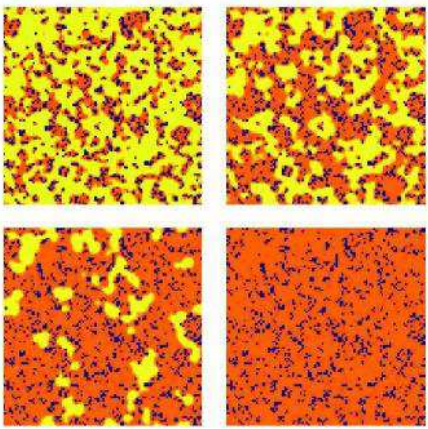}}
\caption{Cross-section of a $87\%$ aerogel sample on adsorption at
$\mu=-5, -4.64, -4.5$ and $-4.35$ (L=100). Gel sites are shown in black
and fluid density is shown as various degrees of grey. As explained in
the text, one can actually distinguishes only two regions corresponding
respectively to $\rho_i\leq 0.05$ (light grey) and $\rho_i\geq 0.95$ (dark grey).}
\end{center}
\end{figure}

In order to better visualize the underlying microscopic mechanism, we
show in Figs. 7 and 8 some cross-sections of typical $87\%$ and $95\%$
porosity samples for different values of the chemical potential along the
isotherms (here, the size of both  systems is $L=100$). One can see 
that in both aerogels  the first stage of the adsorption process is
the formation of a liquid layer that coats the aerogel strands.  Then,
as $\mu$ increases, the film thickens and  condensation occurs in the
smallest crevices defined by neighboring gel strands. In the $87\%$
aerogel, it is difficult to discriminate between these two filling
processes because the available empty space is small, as illustrated by Fig. 4.
  For the same reason, the vapor bubbles that remain  in the system at $\mu=-4.5$
are isolated, and, as they
shrink in size, the adsorption  continues gradually until the solid is completely filled with liquid. This is precisely  the scenario described  in Ref.\cite{TYC1999} from the experimental observations
with $^4$He. This is also similar to what happens in a low-porosity
glass like Vycor\cite{PLAHDW1995}. On the contrary, in the $95\%$
sample, one clearly distinguishes the small capillary condensation events that occur in
some regions of the aerogels (compare the figures for
$\mu=-4.16$ and $\mu=-4.13$) and the major final event that
corresponds to the filling of a large void space spanning the
whole sample (as $\mu$ is increased from $ -4.13$ to $-4.125$). Note
 that the local  ``liquid-vapor'' interfaces inside the
gel are very
sharp at this low temperature, which explains why
one can only distinguishes two different regions in Figs. 7 and 8. Indeed, as
illustrated in Fig. 9 for the lighter aerogel, it is found that
the distribution of the  $\rho_i$'s is bimodal, with
most of the fluid sites having a density lower than $0.05$ or larger than
$0.95$ (the distribution is similar in the $87\%$
porosity aerogel with just a little more intermediate
densities). 

\begin{figure}[t]
\begin{center}
\resizebox{6cm}{!}{\includegraphics{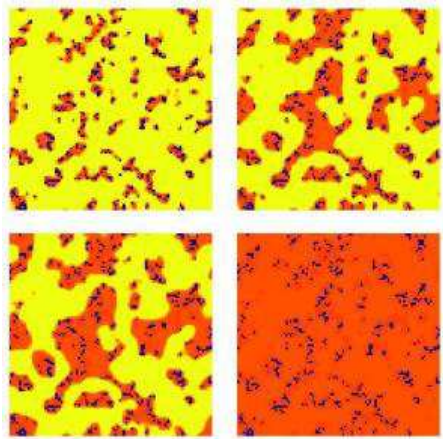}}
\caption{Same as Fig. 7 for  a $95\%$ sample at $\mu=-4.5, -4.16, -4.13$
and $-4.125$ (L=100). This is one of 
the samples shown in Fig.  5.b}
\end{center}
\end{figure}

\begin{figure}[b]
\begin{center}
\resizebox{6cm}{!}{\includegraphics{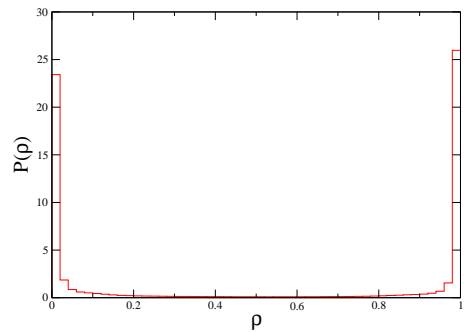}}
\caption{Histogram of the local fluid densities $\rho_i$  in the $95\%$ aerogel sample of Fig. 8 at $\mu=-4.13$.}
\end{center}
\end{figure}
From these results (see for instance the cross-section of the  $95\%$
aerogel at $\mu=-4.13$), there is no indication that the radius of
curvature of the liquid-vapor interface is concave and uniform
throughout the sample just before the major condensation event, as was
suggested in Ref.\cite{TYC1999}. 
This makes unlikely any interpretation in terms of a traditional
capillary condensation model based for instance on the
application of the Kelvin equation\cite{LMPMMC2000,GBBC2003}. Indeed,
condensation in the remaining void space is itself triggered  by a
condensation event that occurs in one (or several) small  region of the aerogel and that induces further
 condensation in the system. 
This collective behavior in the form of avalanches of varying sizes is similar to that of 
 the  Gaussian RFIM at $T=0$ proposed by Sethna and co-workers to
describe the Barkhausen effect in low-T ferromagnetic
materials\cite{S1993}. Increasing slowly the chemical potential of the
adsorbed fluid is equivalent to increasing  adiabatically the applied
field  in a ferromagnet, and changing the porosity of the aerogel
modifies the amount of quenched disorder in the system.  In the RFIM,
the amount of disorder is controlled by the width of the random-field
distribution: only small  avalanches are observed for large disorder,
resulting in a smooth hysteresis loop, whereas  one macroscopic avalanche produces  a discontinuity  in the magnetization for small disorder\cite{S1993,PV2003}.  The transition between these two regimes for a certain value of the disorder corresponds to a critical out-of-equilibrium phase transition at which the distribution of avalanches follows a power law. 

\begin{figure}[t]
\begin{center}
\resizebox{8cm}{!}{\includegraphics{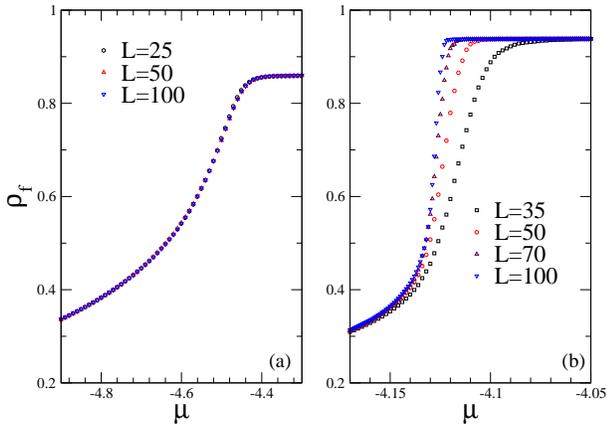}}
\caption{Average adsorption isotherms in $87\%$ (a) and $95\%$ (b)
aerogels for different system sizes. The number of gel realizations
ranges from $500$ ($L=35$) to $20$ ($L=100$) for the $87\%$ aerogel,
and from $2000$ ($L=35$) to $300$ ($L=100$) for the $95\%$ aerogel}
\end{center}
\end{figure}
The existence of a major condensation event in most of the $95\%$
samples of size $L=100$ therefore suggests that there is one macroscopic (i.e., infinite)
avalanche in the thermodynamic limit, corresponding to a finite jump
in the adsorption isotherm.  However, this can be only confirmed by
performing a finite-size scaling analysis of the isotherms obtained
after averaging over many gel realizations. We indeed observe 
significant sample-to-sample fluctuations both in the
location and the height of the largest jump. Such average
isotherms are shown in Fig. 10. For the $87\%$ aerogel, there is obviously
no size-dependence and we can safely conclude that the adsorption is gradual
 in the thermodynamic limit. On the contrary, in the
lighter aerogel, the isotherms  look steeper and steeper
as the system size is increased and the results suggest that there is indeed a
discontinuity when $L\to \infty$ (the isotherm corresponding to $L=35$ is
shown here for completeness, but this size is probably too small to
describe properly a $95\%$ aerogel). As we discussed elsewhere\cite{K2002}
in the case of fluid equilibrium behavior in purely random solids, one 
expects that the maximal slope of the
isotherms should scale as $L^{3/2}$ at a first-order transition
(assuming that the location of the largest jump fluctuates around its mean value $\overline{\mu}_t(L)$ with a
variance $\delta \mu_t(L)^2 \propto L^{-3}$). As shown in Fig. 11, one can obtain a reasonably good collapse of 
the $L=50, 70$ and $100$  irreversible ``compressiblity'' curves $d\rho_f/d\mu$
by using the scaling variable $L^{\chi}\{\mu-
\overline{\mu}_t(L)\}$ with $\chi\approx 1.22$.

There are at least two possible explanations  for the discrepancy  with the expected value  $\chi=3/2$. Firstly,
the system sizes could  still be too small so that too  many gel
realizations would have a non-typical behavior, with several avalanches
of similar heights  (note that in a coarse-grained  picture of a $95\%$
gel-fluid system  where a region  of  size $\xi_G$ is  represented by  a
single effective spin, there would  be only about $10^3$  such
effective spins in a sample  of size
$L=100$). Secondly, one may be  close to the  critical value $p_c$  of the
disorder,  i.e., the critical value  of the porosity for  $y=2$ and $T/T_c=0.45$, for which the
infinite avalanche first  appears.  Studies  of the  $T=0$
RFIM  have  moreover shown that  the  critical  region  is unusually
large\cite{S1993}. The value $\chi=1.22$ is compatible with the
predictions of Sethna and co-workers, with $\chi=\beta\delta/\nu\simeq2-\eta$\cite{note5}, although one cannot also exclude that 
the exponents differ from those of the conventional RFIM
because of the presence of impurities\cite{T1996}.
In any case, it seems reasonable, on the basis of the present
calculations, to predict  the existence of a phase transition with
most probably a
discontinuity in the adsorption isotherm in the thermodynamic limit,
as shown in Fig. 5(b).

\begin{figure}[t]
\begin{center}
\resizebox{8cm}{!}{\includegraphics{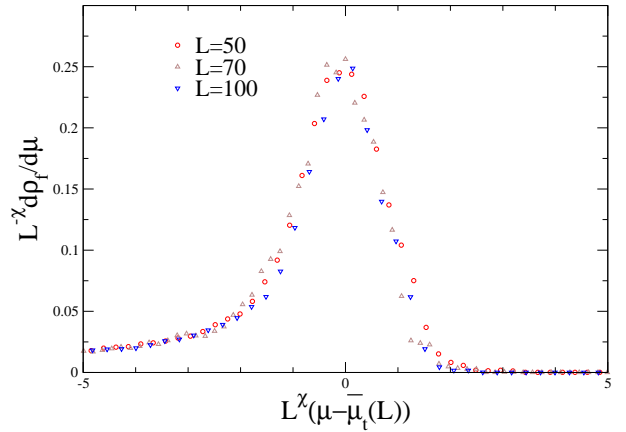}}
\caption{Scaling plot of the compressibility curves $d\rho_f/d\mu$  during adsorption in
the $95\%$
aerogel with $\chi=1.22$ ($\overline{\mu}_t(L)=-4.123, -4.126, -4.128$ for $L=50, 70,
100$, respectively; extrapolation to $L\to \infty$ gives $\mu_t=$lim$_{L\to \infty}\overline{\mu}_t(L)\simeq -4.131$).}
\end{center}
\end{figure}

\subsection{Desorption }
\begin{figure}[t]
\begin{center}
\resizebox{8cm}{!}{\includegraphics{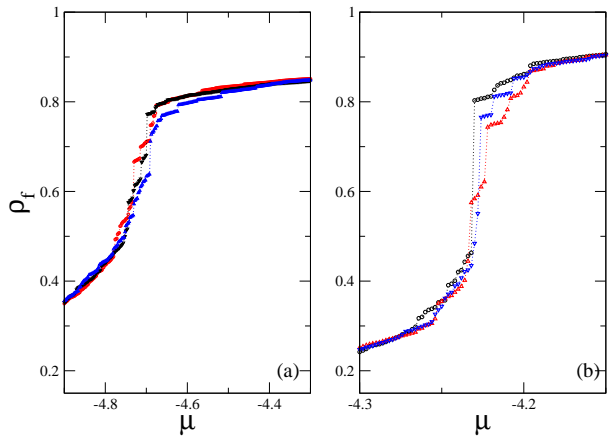}}
\caption{Desorption isotherms in $87\%$ (a) and $95\%$ (b)
aerogels. System sizes are $L=50$ (a) and $L=100$ (b).}
\end{center}
\end{figure}

As discussed elsewhere\cite{R2003}, fluid desorption in disordered
porous solids can take place via several different 
mechanisms, depending on the temperature and on the  
structural and energetic properties of
the solid (in Ref.\cite{R2003}, however, only the influence
of the interaction parameter $y$, i.e., of the wetting properties of
the adsorbed fluid, was described). In
a first mechanism, desorption is due to the appearance of vapor
bubbles  in the bulk of the material, bubbles that grow, coalesce, and
eventually extend over the whole pore space. The mass adsorbed 
then decreases continuously. In
the other mechanisms, draining of the solid starts from the surface, and the 
desorption is associated to the penetration  of a
vapor-liquid interface which was previously pinned by the
irregularities of the solid structure for $\mu$ larger than some threshold
value $\mu_c$ (the ``depinning''threshold). The desorption curve is then
 either gradual or discontinuous,
depending on whether the growth of the vapor domain is isotropic and
percolation-like (self-similar), or compact (self-affine). In those cases the transition is  however always critical because the scale of the rearrangements of the interface diverges as  $\mu \to \mu_c^+$.  This 
is very similar to the physics  of fluid invasion in porous media\cite{CR1988}, although we are considering here a single compressible fluid, and of field-driven domain wall motion in disordered magnets\cite{KR2000}. 

 In order to find what are the relevant mechanisms in the
$87\%$ and $95\%$ aerogels for $y=2$ and $T/T_c=0.45$, we have 
studied the two systems
in the presence and in the absence of the interface with the gas reservoir. 
In the latter case, we find that emergence of vapor bubbles in the bulk of
the solid only occurs  at low values of the chemical potential ($\mu
\simeq -5.27$ and $\mu\simeq -5.25$ for $87\%$ and $95\%$ aerogels, respectively),
very close to the value $\mu_{spi}=-5.249$ corresponding to the liquid  (mean-field) spinodal
of the bulk fluid at $T/T_c=0.45$. This shows that the perturbation induced by the
solid is too small to displace significantly the liquid spinodal. On
the other hand, the contact with the ambiant vapor at the surface of the gel has a
major influence, as illustrated in Fig. 12 by typical desorption
isotherms  calculated  in the presence of a free surface. One can see that when decreasing  the chemical potential 
all the curves exhibit a pronounced drop  much before vapor
bubbles appear in the bulk of the material. This  is a clear
indication that the mechanism of desorption is due to the surface.

\begin{figure} [t]
\begin{center}
\resizebox{6cm}{!}{\includegraphics{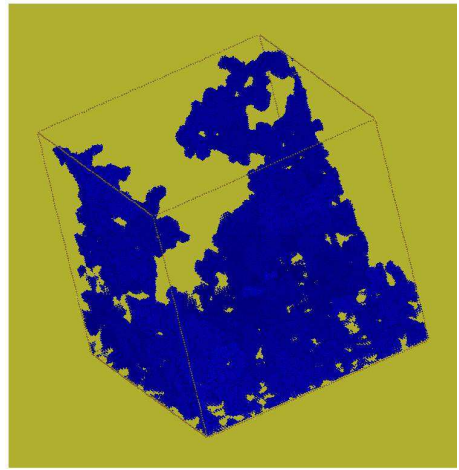}}
\caption{Invading vapor domain $V_t$ (in dark) in a $87\%$
aerogel at $\mu=-4.706$ ($L=100$). $10\%$ of the fluid has drained out. The gas
reservoir located at the bottom of the box and the aerogel are not shown.}
\end{center}
\end{figure}

\begin{figure}[b]
\begin{center}
\resizebox{6cm}{!}{\includegraphics{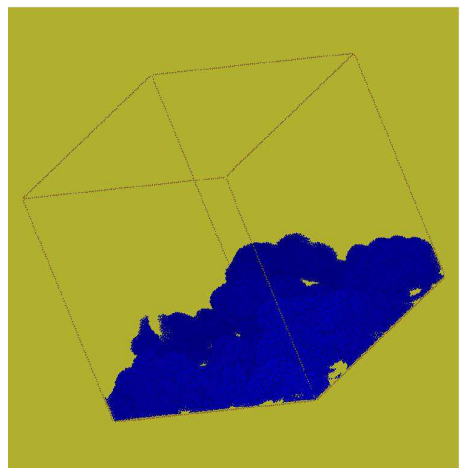}}
\caption{Same as Fig. 13 but for a $95\%$ aerogel at $\mu=-4.220$. }

\end{center}
\end{figure}

The isotherms shown in Fig. 12 consist of many steps of varying size,
but it also appears that one step is significantly
larger than the other ones in most of the $95\%$ samples,
a feature that is not present in the isotherms of the $87\%$ aerogel.
This suggests that the growth morphology of the invading
vapor domain may change with the porosity. This is confirmed by the snapshots
displayed in Figs. 13 and 14 that show the invading vapor domain 
in two typical $87\%$ and  $95\%$ samples when about $10\%$ of the fluid has
drained out of the aerogel. 
This is just after the onset of the sharp drop in the isotherms. As in the case of adsorption, we find that
the distribution of the  $\rho_i$'s is bimodal in
this range of chemical potentials, with
most of the fluid sites having a density lower than $0.05$ or larger than
$0.95$. One can thus identify unambiguously the emptied
sites. Although the volume occupied by the vapor is
the same in the two samples (about $200000$ sites), the morphology of the domain is remarkably
different. In the $87\%$ aerogel, the vapor domain exhibits an intricate isotropic
structure that resembles that produced in
invasion percolation. In contrast, the domain looks compact with a self-affine
interface  in the lighter aerogel. This is very similar to the two regimes observed in the $T=0$ RFIM when an  interface separating  two magnetic domains  is driven by an external field\cite{JR1992}:  when the disorder is large, the interface forms a self-similar pattern with a large-scale structure characteristic of percolation, whereas the growth is compact and the domain wall forms a self-affine fractal  surface at intermediate degrees of disorder (note that the use of the body-centered cubic lattice has allowed us to suppress the faceted growth regime that is observed at $T/T_c=0.45$ in the  $95\%$ porosity aerogel on the simple cubic lattice\cite{note6}; this regime is  an artifact of the lattice description).

In order to confirm the existence of two growth regimes and to determine
the actual behavior in the thermodynamic limit, a finite-size scaling
analysis of the desorption isotherms is required.  Average isotherms
calculated for different system sizes are shown in Fig. 15. By analogy
with the problem of a driven interface  in the $T=0$ RFIM, one expects
that  the total volume $V_t(\mu)$ of the invading vapor domain shows a
power law divergence at the depinning threshold $\mu_c$. Then, assuming
that  the only relevant length scales  near  $\mu_c$ are the sytem size
$\L$ and a single correlation length $\xi\sim[(\mu -\mu_c)/\mu_c]^{-\nu}$, the
dependence of $V_t$ on sytem size should be described by the scaling
form  $L^{D_f}g(x)$ where  $D_f$ is the fractal dimension
characterizing the domain and $g$ is a universal function of the
scaling variable $x=L^{1/ \nu}( \mu -\mu_c)/\mu_c$\cite{JR1992}.
 
In both aerogels,  one observes important  finite-size
effects (see Fig. 15), but, unfortunately, they are not only due to the existence of a diverging
correlation length in the system but also to boundary effects. There is indeed an initial regime in which 
desorption is due to the draining of large crevices at the surface of the gel  where the fluid is in direct
contact with the ambiant vapor. For a given porosity, the number of
these crevices is proportional to $L^2$ and the contribution to the fluid
density is thus proportional to  $1/L$. It turns out that this initial regime
extends to rather low values of the chemical potential ($\mu \simeq -4.6$ and
$\mu \simeq -4.15$ for the $87\%$ and $95\%$ aerogels, respectively), so that
the scaling region around the depinning threshold is too small to be
studied properly  and to extract the  values of the critical exponents (such an effect is not present 
in the numerical studies of domain growth in the standard $T=0$
RFIM\cite{KR2000} because there is a different random field on {\it
each} site of the lattice and no equivalent of the crevices). We
note, however, that the curves in Fig. 15(b) have a common
intersection at $\mu\simeq -4.24$, in contrast with those in Fig. 15(a). This  is
consistent with the scaling ansatz for the volume of the invading
vapor domain with $D_f=3$ for the $95\%$ aerogel and
$D_f<3$ for the $87\%$ aerogel. We thus conclude that the 
desorption seems to be discontinuous in the first case and gradual
(percolation-like) in the second one. 

This leads to the isotherms shown in Fig. 5. Their shape resembles that of the experimental curves
in Figs. 4(b) and 4(c) of Ref.\cite{TYC1999}. The minor differences
can be rationalized:
the experimental isotherm in the $87\%$ aerogel (Fig. 4(c) in
Ref.\cite{TYC1999}) does not exhibit a
sharp kink at $\mu_c$, but this rounding may be due to the activated
processes that are neglected in the present treatment; 
Fig. 4(b) in Ref.\cite{TYC1999}  corresponds to a $98\%$ aerogel, which
probably explains why the shape of the hysteresis is more rectangular
than in the present Fig. 5(b).

\begin{figure}[t]
\begin{center}
\resizebox{8cm}{!}{\includegraphics{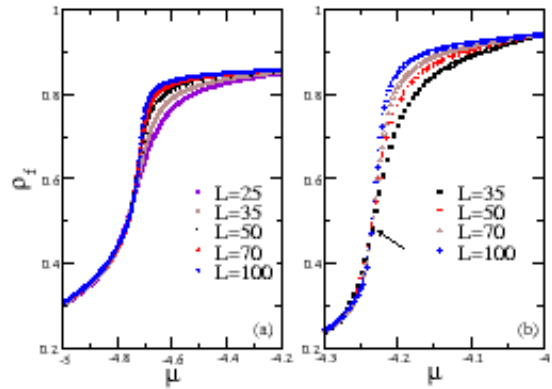}}
\caption{Average desorption isotherms in $87\%$ (a) and $95\%$ (b) aerogels for different system sizes. The number of gel realizations
ranges from $1000$ ($L=25$) to $100$ ($L=100$) for the $87\%$ aerogel,
and from $1000$ ($L=35$) to $200$ ($L=100$) for the $95\%$ aerogel. The
arrow in (b) indicates the common intersection of the curves.}
\end{center}
\end{figure}

\subsection{Equilibrium }

In order to determine the equilibrium isotherms, one has to find,
for each value of $\mu$, the lowest lying state(s) among all the
metastable states obtained from 
Eqs. (3). A complete enumeration of these states is, however, an
impossible numerical task, and like in previous
work\cite{K2001,K2002}, we have only 
calculated  a limited number of states that, hopefully, can
provide a good approximation of the equilibrium isotherm. This can be
checked {\it a posteriori} by using the Gibbs adsorption equation
$\rho_f=-\partial(\Omega/N)/\partial\mu$ which is only satisfied by the equilibrium
curve\cite{K2001}. In Refs.\cite{K2001,K2002}, we searched for metastable states inside the
hysteresis loop by starting the iterative procedure with initial
configurations corresponding to uniform fillings of the lattice (with $\rho_i^{(0)}=\rho$
varying between $0$ and $1-p$). This method, however, does not
converge in dilute aerogels because the structure is very
inhomogeneous and correlated, at least for distances smaller than
$\xi_G$. It is then likely that all  metastable fluid configurations are
also very inhomogeneous  and cannot be reached  iteratively from
initial uniform fillings. 
We have thus searched for metastable states by calculating desorption and adsorption scanning curves, i.e., by performing incomplete filling or draining of the aerogel and then reversing the sign of the evolution of the chemical potential.
For convenience, these (very long) calculations are done
in systems with  periodic boundary conditions in all directions, i.e.,
in the absence of an interface with the reservoir\cite{note7}.

\begin{figure}[t]
\begin{center}
\resizebox{8cm}{!}{\includegraphics{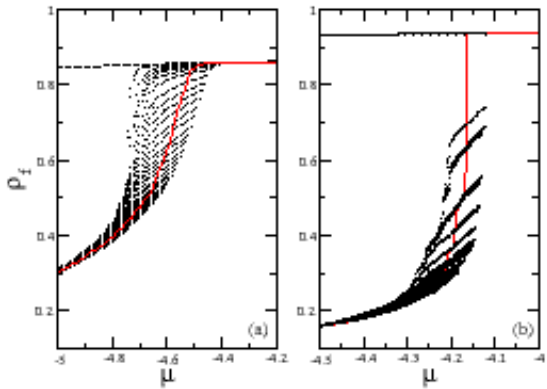}}
\caption{Typical desorption scanning curves (points) and equilibrium isotherms
(solid lines)  in  $87\%$ (a) and $95\%$ (b) aerogels (L=70).}
\end{center}
\end{figure}

Some typical desorption scanning curves are shown in Figs. 16(a) and
(b). In both cases, the top curve is the major desorption branch, and  by
comparing with the isotherms shown in Figs. 12 or 15, it is clear that
the part of this branch extending to low $\mu$'s and corresponding to
liquid-like metastable states that are isolated from the other states
is an artifact coming from the absence of the interface with the
reservoir. We also notice that  there are no scanning curves in the
upper part of the hysteresis loop for the $95\%$ sample. This is
because there is a jump in the adsorption isotherm and, thus, there are  no intermediate states from which 
desorption can be started. Therefore the states contained in  this region of the
plane $(\rho_f,\mu)$ cannot be reached by performing scanning
trajectories\cite{note8}. (Further work is clearly needed to determine if there are no metastable
states at all in this region or if the states can be obtained by other
means, for instance by changing the temperature.) 

\begin{figure}[t]
\begin{center}
\resizebox{8cm}{!}{\includegraphics*{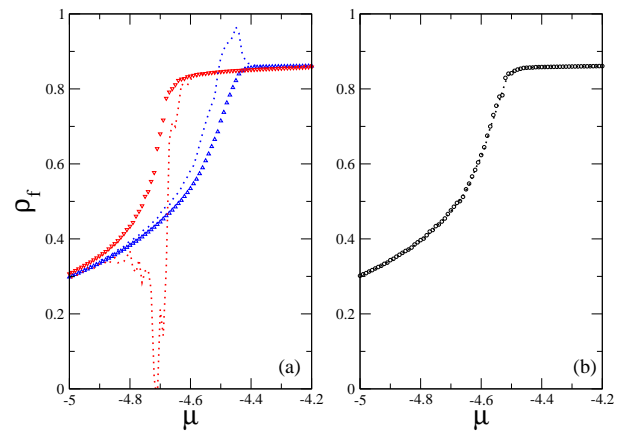}}
\caption{Check of thermodynamic consistency along the adsorption,
desorption, and equilibrium isotherms for a $87\%$ porosity aerogel
($L=70$). (a) adsorption and desorption (b) equilibrium. Symbols:
average fluid density $\rho_f$. Dashed curves: related quantity obtained
by differentiating the corresponding grand potential with respect to
the chemical potential.}
\end{center}
\end{figure}

\begin{figure}[b]
\begin{center}
\resizebox{8cm}{!}{\includegraphics*{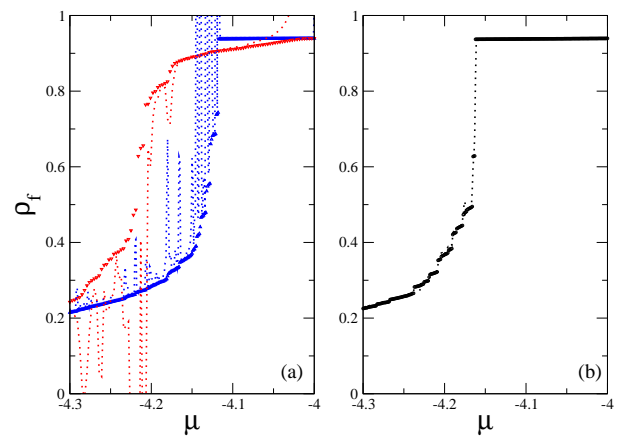}}
\caption{Same as Fig. 17 but for a $95\%$ aerogel.}
\end{center}
\end{figure}

Fig. 16 also shows the approximate equilibrium isotherms obtained  by
selecting, for each value of $\mu$, the state $\alpha$ that gives the lowest value of the grand
potential, as calculated from Eq. (4). We have checked that taking into
account the additional metastable states 
obtained from the adsorption scanning curves did
not change significantly the results (as in Ref.\cite{K2002}, we have also
checked that keeping all solutions with a weighting factor equal to the  
Boltzmann factor gives the same isotherms). As illustrated
in Figs. 17 and 18, the Gibbs adsorption equation is very well
verified along these curves. This indicates that we have indeed obtained  a good
approximation of the true equilibrium isotherms. In contrast, 
thermodynamic consistency is strongly violated along the adsorption
and desorption branches. The presence of  (delta) peaks
 in $\partial\Omega/\partial \mu$ also shows that the grand potential $\Omega$  changes 
discontinuously (as the fluid density $\rho_f$) during adsorption and desorption in 
a single finite sample. On the other hand,  $\Omega$ is continuous (but
$\rho_f$ is discontinuous) along the equilibrium isotherm, as already noticed in Ref.\cite{K2002}.

One can see from Figs. 16(a) and 16(b) that the  equilibrium behavior
is different in the $87\%$ and $95\%$ porosity aerogels.
 In particular, there is a large final jump in the isotherm
of  Fig. 16(b). The same feature
exists in all samples, but in order to conclude on the actual behavior
in the infinite system one has again to analyze the size-dependence of
the average isotherms. This is shown in Fig. 19. For the $87\%$ aerogel, there is almost no
 size-dependence and it is clear that no
transition occurs when $L\to \infty$. On the other hand, for
 the  $95\%$ aerogel, the isotherms become steeper as $L$ increases and there is
a rather well defined common intersection at $\mu_t\simeq -4.17$. This is a
clear indication that a phase transition does occur in the
thermodynamic limit. However, we have not
succeeded in obtaining  a satisfactory collapse of the
isotherms using the scaling reduced variable
$L^{3/2}[\mu-\mu_t(L)]/\mu_t(L)$ as was done in Ref.\cite{K2002} in the case
of a random solid. The origin of this
problem is still unclear, but we suspect that the system sizes  are
again too small. Indeed, we note  that the maximal slope increases
rather weakly between  $L=35$ and $L=70$, but has a size-dependence
consistent with the exponent $3/2$ between $L=70$ and
$L=100$. We thus conclude that our results are consistent with a discontinuous jump in
the thermodynamic limit, as indicated in Fig. 5(b). This corresponds
to a true equilibrium liquid-vapor phase separation in the system.

\begin{figure}[t]
\begin{center}
\resizebox{8cm}{!}{\includegraphics{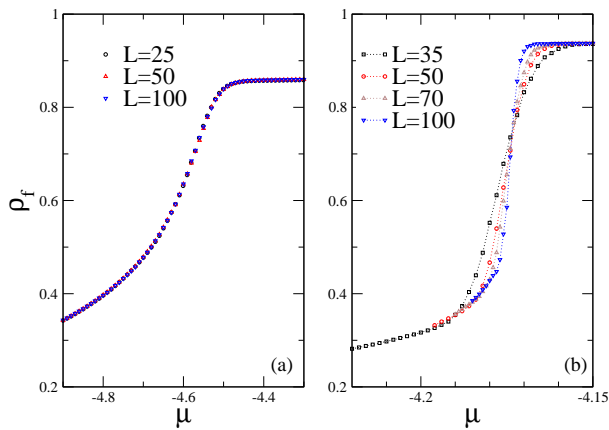}}
\caption{Average equilibrium isotherms  in  $87\%$ (a) and $95\%$ (b) aerogels for different system sizes. The number of gel realizations
ranges from $500$ ($L=25$) to $20$ ($L=100$) for the $87\%$ aerogel,
and from $2000$ ($L=35$) to $200$ ($L=100$) for the $95\%$ aerogel.}
\end{center}
\end{figure}

\section{Summary and conclusion} 

In this paper, we have proposed an interpretation of the hysteretic
behavior observed in the experiments of $^4$He adsorption in light silica
aerogels. The overall shape of the experimental hysteresis
loops is well described by our theoretical model and we have been
able to reproduce the dramatic influence of porosity.  In this interpretation, the disordered character of 
the aerogel structure plays an essential role, the porosity $p$ being the tunable parameter that controls 
the amount of disorder in the system. The history-dependent behavior is thus associated to the presence of
many metastable states in which the system may be trapped and that prevent thermal equilibration
at low temperatures. The most
important  conclusion of our study is that adsorption and desorption
obey to different mechanisms and may be gradual or discontinuous,
depending on the porosity. Adsorption is insensitive to the presence of the interface between the solid and the 
external vapor, and the change in the shape of the isotherm from smooth
to discontinuous as $p$ increases from $87\%$ to $95\%$ has been related to the appearance
of a infinite avalanche occuring in the bulk of the system, similar to
what happens in the $T=0$ RFIM below the critical
disorder\cite{S1993}. In contrast, desorption 
is triggered  by the presence of the outer surface and an associated
depinning transition. The change in the shape of the
isotherm is then related to a change in the morphology of the
invading vapor domain from percolation-like to compact (note however 
that the mechanism
may be different in aerogels of lower porosity\cite{D2003}).

One must keep in mind that our calculations are based on local-mean
field theory in which disorder-induced fluctuations are properly
accounted for but thermal
fluctuations are neglected. This appears to be a reasonable assumption
at very low temperatures where the time scale associated to thermally
activated processes is much larger than the experimental time
scale. In this case, both the adsorption and desorption branches are
metastable and we have shown that the true equilibrium isotherm
(which probably cannot be reached experimentally) is 
somewhere in between: this isotherm also changes from gradual to
discontinuous as $p$ increases.  The situation is different at higher temperature (in
the vicinity of $T_c$) since true equilibrium behavior without
hysteresis has been
observed experimentally\cite{WC1990,WKGC1993}. 
It will be therefore of interest to study the influence of temperature
on the hysteretic behavior  and to understand how the dynamics of relaxation
towards equilibrium changes with $T$ (see, e.g., Ref.\cite{WM2003} for
a recent study of the relaxation behavior associated to  capillary
condensation in a lattice model of Vycor).

Finally, we suggest that the scenario for filling and draining 
described in this work could be tested in more detail by
performing experiments in a series of aerogels of gradually increased
porosity.  If our interpretation of the adsorption process in
terms of avalanches is correct, there should exist a value of
the porosity for which the isotherm becomes critical and the
distribution of avalanche sizes follows a power law with well-defined
critical exponents\cite{S1993}. Using superfluid instead of normal
$^4$He could perhaps allow to observe distinct avalanche events in
the aerogel and to study their statistical properties. Such a study has
been performed recently in the nanoporous material  Nucleopore by a
capacitance technique\cite{LH2001}. One could also
check that the draining of the aerogel starts from the surface and
that the growth of the invading vapor domain obeys different
regimes. Related studies have been for instance  carried out in Vycor using
ultrasonic attenuation and scattering
techniques\cite{PLAHDW1995,K2000}. Such experiments would give a
definite answer to the long-standing question about the nature of
hysteresis in fluid adsorption in disordered porous media.

\acknowledgments
The Laboratoire de Physique Th\'eorique des Liquides is the UMR 7600 of
the CNRS. We thank R. Jullien for providing us with his lattice DLCA algorithm
to build the model aerogel.

	

\begin{thebibliography}{10}
\bibitem{note3} We consider here  hysteretic cycles  that are observed
at very slow flow rates on filling and on draining when the system
exhibits no explicit time-dependence (after some transient time). This corresponds to the so-called rate-independent  limit.
\bibitem{note6} In the absence of gel, there is no 
threshold for the onset of motion of the initial liquid-vapor
interface parallel to the $[100]$ plane of the bcc lattice, i.e., the interface moves
along the $x$-axis for $\mu<\mu_{sat}=-4$ at all temperatures. The actual
self-affine interface, however, has various orientations at short-length scale.
We found  numerically that, at $T/T_c=0.45$, a portion of the 
interface parallel to the $[110]$ plane starts to
move at $\mu= -4.11$. Since the 
drop in the desorption isotherms of Fig. 12  occurs at lower
values of $\mu$, the pinning of the interface is due to disorder and not
to lattice effects.
\bibitem{note7} In the thermodynamic limit, the boundary condition
(presence or absence of an interface) is not expected to influence the
equilibrium phase behavior in the bulk of the material. As we have seen
before, this reasoning does not apply to metastable states and
out-of-equilibrium behavior.
\bibitem{note8} The presence of a jump in the adsorption isotherm also
implies that there are no  adsorption
scanning curves in the upper part of the hysteresis loop. Indeed, in
the absence of interface with the
reservoir, these curves are obtained by re-adsorbing from the
metastable states located along the last desorption scanning curve.
\end{thebibliography}
\end{document}